# Can Research be Taught?

**Mouhamed Abdulla and Yousef R. Shayan**
Department of Electrical and Computer Engineering
Concordia University
Montréal, Québec, Canada
Email: {m_abdull, yshayan}@ece.concordia.ca

## Abstract

*The word "researcher" is loaded and often confusing. It takes years to become one and to master all of its aspects. In this paper, we investigate whether or not this process of "becoming" can be catalyzed through education. The focus will be on wireless communications, though the same principle could very well be replicated to other disciplines.*

## 1 Introduction

Throughout the world, undergraduate studies are usually based on courses; thus, the structure for graduation is straightforward and for the most part predictable. After graduation, once a high-achieving student decides to pursue graduate work, they are usually interested by what the degree being pursued symbolizes and how it may move them forward in their carrier path. It turns out, from several case observations that because the principle focus was on the reward, as oppose to the mean that ensures a successful and on time graduation, important challenges might occur.

Given this sudden jump from an organized and deterministic undergraduate curriculum to a more irregular and non-linear thesis based work, most students will pass through a difficult phase of shock and surprise. In fact, because of such reasons, among others, after the minimum required coursework is completed, many may decide to discontinue the program. Others, with greater motivation, perseverance and defiance will try to face these obstacles to the best of their ability. Usually, credible universities have counselling offices that can assist graduates in such difficult phases. Another segment of students, will try to seek direct guidance from their advisor, and sometimes from senior peers.

Obviously, thesis work is based on original contribution(s). And, in terms, contribution is a by-product of research. Hence, in this paper, as the title suggests, we will try to carefully build a case and give an answer to the question of whether research can be taught. On one hand, original discovery is expected, which requires talent. However, such an aptitude is not necessarily acquired because of indoctrination fears, and also it is expected to be intrinsic or inherent. On the other hand, there are worthwhile common denominators for design-based research that can indeed be learned.

Clearly, the motivation of this paper is philosophical in nature and esoteric in depth. Yet, we will try to identify, demystify and explore what research entails. This will be done, by providing clear guidelines and effective performance benchmarks that can be used to prepare and assist engineering graduate students throughout their program.

The treatment will be divided into several sections. First, from an engineering point of view, we will attempt to provide a practical significance to the word *research*. Next, we will outline the essential prerequisites required before the start of a new project. Then, we will show a tested, coherent, and systematic framework that can facilitate a genuine research idea and direction. Further, we will elaborate on the most effective ways to perform academic search. Finally, we will explain the metric needed to measure the worthiness of a research principle. All the above will be done from a telecommunications and wireless engineering perspective, though the approach may very well be extended to other design-based fields.

## 2 What is Research?

Ultimately, the essence of this paper is to try to understand how we can effectively *build* a researcher. However, before we get into this, it is imperative to first understand what research really suggests.

In fact, if a random sample of society is surveyed, it is almost certain that a large number of definitions to the term research will emerge; and some of these are listed in [1]. Nonetheless, the meaning that most might probably agree on is that research is the action carried during information gathering.

This, from a student's perspective, could translate to reading, collecting data of interest, and writing a report. While such an interpretation is an important step toward genuine discovery because it acts as a literature review or survey, though as a standalone approach, it does not define academic graduate-level research. The reason is that it fails to provide novel contributions to the already available body of knowledge. Hence, simply put, we define engineering research as *the uncovering of an original, practical and efficient: principle, model, or gadget.*

To make this more tangible, Figure 1 shows our understanding of how original research is produced. In particular, we start by asking a question that intrigues and interest us. Next, we go to the public domain and search for available knowledge hosted by credible and well respected publications. If an answer to our question is found, then we try to further push the limits by figuring out if we can improve the answer. If we can, it then implies that we have found our research direction. If on the other hand we cannot improve the answer, we loop-back and ask yet another question, which could very well be adjacent to the original interrogation.

However, if we establish, to the best of our ability, after careful investigation, that an answer to our question is not available in the public domain, then we hypothesize a possible way to solve the problem. It should be clear that hypothesis is simply a guess. And this guess could be trivial, nontrivial, or it could be based on other studies somewhat relevant to the orientation of the proposed question, etc.

On a side note, it is interesting to mention that the famous Italian/American physicist Enrico Fermi once said: *"there are two possible outcomes: if the result confirms the hypothesis, then you've made a measurement. If the result is contrary to the hypothesis, then you've made a discovery"* [2]. So following this logic, in either case a conclusion can be made, which is what we are after.

In spite of this, if we do not have any clue of what the hypothesis could be, we could then introduce assumptions such that the original question becomes simpler, and then repeat the process. Also, one needs to be careful about the differentiation between hypothesis and assumptions; they are not the same, as detailed in [1].

Therefore, to sum-up this section, in our opinion novel engineering research is one of two things: *Either we can improve a known idea, or we have a new idea all together.* And this is coherently illustrated in the flowchart of Figure 1.

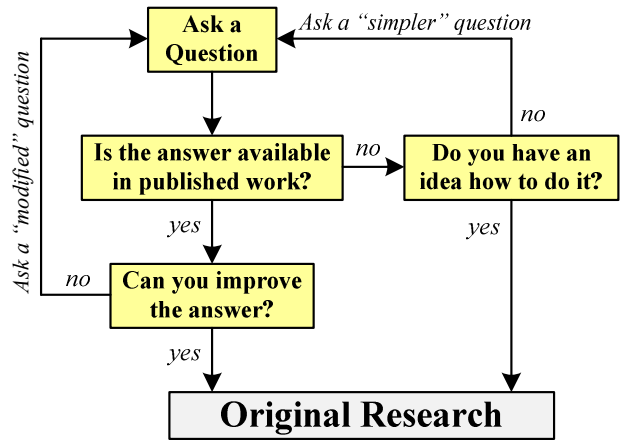

Figure 1. Research flowchart.

## 3 The Prerequisites for Research

One cannot start researching a topic of interest right away, there are specific technical prerequisites needed. This is why, for the most part, research activity starts only after graduation, while enrolled in a master's or a doctorate program.

For instance, in PHYsical (PHY) layered wireless telecommunications, students are expected to have a strong engineer mathematical background. At the same time, it is hoped that these students, interested in graduate-level telecom research, have at least been exposed to the fundamentals of signal processing. Clearly, the more the better, though these two courses are the very minimum qualifications desired.

Besides, for the most part, the student's advisor will recommend the courses to take as a function of the expected research direction. This approach is excellent, provided the direction is known *a priori*. However, this is not always the case, and hence generic courses, from a PHY perspective, are then suggested as in the example of Figure 2.

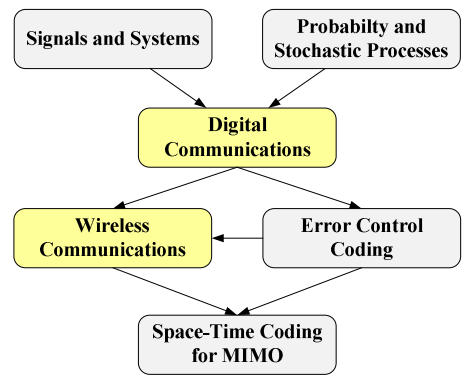

Figure 2. PHY wireless communication courses.

In fact, in the illustration shown above, and perhaps obvious from their proper interdependencies, it should be apparent that the foundation for research at the system-level is strongly dependent on the digital and wireless communication courses. In particular, the digital course will introduce the concepts and fundamentals of transmission from a guided-propagation point of view (i.e. wireline). Then, the wireless course will pick what was learned in the previous semester and complement it by the effect of free-propagation (i.e. wireless). The breakdown of these steps was described in our previous paper [3].

## 4 Getting a Research Idea

This may seem exceptional, but we have noticed that if a student is capable to have a global picture of the research field in which he or she is involved, then it will effectively help them in properly selecting a specific topic of focus.

Essentially, this method could be interpreted as a *funnel*-like approach that starts wide and ends narrow. To bring this principle closer to mind, let us once again consider the wireless engineering field.

First, we try to find a classification on how communications is divided. In telecom, a known categorization is that of the Open System Interconnection (OSI) model which was created by the International Organization for Standardization back in 1977. Notice in Figure 3, that the model is divided into two layers: upper and lower. In general, engineers specialize in one of them, but not both. For us, our scope is based on the lower part, specifically at the PHY section; though we might also look at issues related to the MAC and Network for cross-layer design and optimization. And if searched, it is almost certain that similar dichotomies are also available or possible in a diverse host of research fields.

Next, we try to list all active wireless standards available in the market. And, we do this, because as mentioned in Section 2, we want to come-up with a *practical* research question that can produce instant return, and will directly affect the well-being and progress of the society based on current needs. However, it should be stressed that research not linked to a specific system still exist. Yet, it is often forgotten because its applications, benefits, and usefulness are *too theoretical* and not immediately obvious or apparent. An interesting example of this is that of Low Density Parity Check (LDPC) Codes invented in 1963 by the well respected MIT Professor Robert Gallager as part of his doctorate thesis [4]. His contribution was completely forgotten until 1996 where Professor David MacKay, a physicist from university of Cambridge rediscovered it [5].

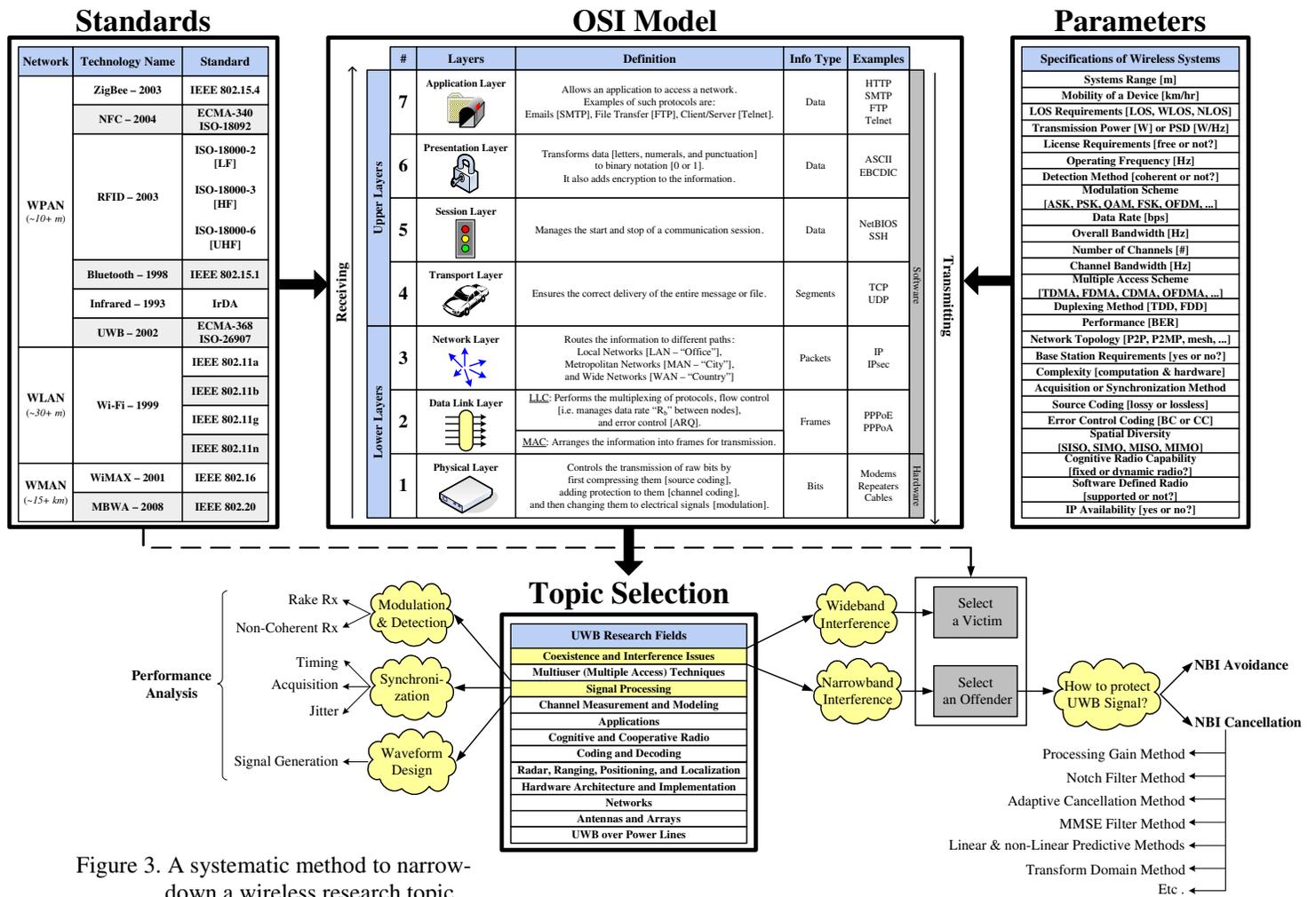

Figure 3. A systematic method to narrow-down a wireless research topic.

Going back to the model of Figure 3, we then list all parameters that could be varied or studied of relevance to telecommunications. In terms, these are specifications that we always hear, see, and read in conferences, presentations, books, papers, graphs, charts, tables, etc.

Now, based on this big picture approach, a topic, a standard, or a system is selected for research. In our example, this topic is called Ultra Wideband (UWB). Next, all interesting subjects within UWB are outlined. And so on and so forth, we continue along a similar approach until we narrow-down the research topic of interest. In Figure 3, two research direction examples are described: coexistence and signal processing.

Admittedly, we find this method interesting because it keeps graduate students grounded to practical real-life design applications, yet at the same time it promotes the focus needed on a very narrow subject. The other benefit of this approach is that it connects the various fields' together; as a result, new research directions and extensions become simpler, clearer and more evident. Moreover, it can bridge the gap during research collaborations because the cause and effect of parameters can more effectively be observed and understood.

On a parallel note, to find a research question, it is also wise to monitor what peers are working on. One of the ways to do this would be to keep up-to-date with new publications. Also, in a given month, it is particularly stimulating to remark what literature is most popular among researchers. To do this, one could for instance look at the Top-100 documents, covering all sponsored topics accessed on IEEE Xplore [6]. Likewise, the IEEE Communications Society website maintains a list of the Top-10 most viewed articles [7].

## 5 How to Search?

The principle of digital link started back in 1948 by Shannon's revolutionary paper published in the Bell System Technical Journal called "A Mathematical Theory of Communications". Since then, until now, so many theories and ideas have emerged. As a consequence, serious new researchers would need to get up to speed on important milestones of the past few decades. Indeed, this task would prove to be challenging because of the incredible and undeniable expansion to the body of research. But at the same time, the technology in searching materials has amazingly improved. The availability of search engines has changed the way we gather and look for information with an unbelievable response time.

This indicates that today's graduate students could get information ubiquitously and quickly. In fact, predominantly, the databases of IEEE Xplore [8], ACM Portal [9], SpringerLink [10], Elsevier ScienceDirect [11], and SIAM Journals Online [12] are some of the best search tools for telecom researchers, as well as others. As for full copy electronic access of previous theses, ProQuest [13] is the mechanism that could be used to get most dissertations, since 1861 up until now. Furthermore, instead of looking at mentioned tools as discrete elements, as a motivation to enhance time/effort efficiency, one can use the MetaFind program [14] capable to inquire all the above with a simple input.

Moreover, in the province of Québec, we have COLOMBO's Inter-Library Loans program [15], sponsored by CREPUQ [16]. This web-application enables the delivery, free of charge, of books not available in the student's local library, from any university within the province, and sometimes abroad.

## 6 A Metric for Research

To measure one's research merit and worthiness is based on how others, who are specialists in the field, think of the proposed work. In essence, the peers are the body that will make the final judgment on the acceptance or rejection of a contribution.

In general, some of the known ways to communicate genuine results are based on peer-reviewed conference proceedings, journals, magazines, seminars, or even panel discussions. In particular, attention should be devoted to publications because it will last forever.

While publishing, one needs to work on important research problems [17], such that hopefully it remains relevant for the next 10 to 20 years, or beyond. In other words, we should publish work that will force others to cite it for a *long-time*. And this, by itself, is an important testament to a contribution.

For clarification, Table 1 shows a sample of some of the most important publications in wireless,

networking, and modeling research. As it can be observed, usually the merit of a conference is measure by its Acceptance Rate (AR). On the other hand, for journals, the Impact Factor (IF), which accounts for the number of times an article is cited [18], is the metric commonly used to quantify and compare a publication.

Table 1. Example of publication metrics for telecom.

| Conference Proceedings | | Journals and Magazines – 2006/2007 | |
| --- | --- | --- | --- |
| Names | AR | Names | IF |
| *InfoCom* | *~ 20 %* | *IEEE/ACM Transactions on Networking* | *1.789 ǀ 1.831* |
| *MSWiM* | *20 ~ 25 %* | *IEEE Journal on Selected Areas in Communications* | *1.816 ǀ 1.799* |
| *ICC* | *~ 35 %* | | |
| *CAMAD* | *~ 35 %* | *IEEE Communications Magazine* | *1.678 ǀ 1.704* |
| *MobiWac* | *~ 35 %* | *IEEE Transactions on Network and Service Management* | *2.211 ǀ 1.609* |
| *CCNC* | *~ 35 %* | | |
| *GLOBECOM* | *35 ~ 40 %* | *IEEE Transactions on Communications* | *1.208 ǀ 1.302* |
| *PIMRC* | *~ 43 %* | *IEEE Transactions on Wireless Communications* | *1.184 ǀ 1.234* |
| *WCNC* | *~ 45 %* | | |
| *VTC* | *~ 45 %* | *IEEE Communications Letters* | *0.684 ǀ 0.869* |

While "AR" and "IF" measures are commonly used by most publications, they do not necessarily echo the quality of a research contribution. In words, the approach is not perfect and clearly has lots of flaws, misuses, and biases. This is why an effort has been made to create an International Symposium on Peer Reviewing (ISPR) that analyses publication activities. In short, their objective is to *"apply peer review to current peer reviewing methodologies"* [19].

## 7 Conclusion

The aim of this paper, though somewhat standard to senior well established researchers, was to show how effectively, coherently, and systematically the philosophical concept of engineering research could be taught to newly admitted graduate students, so as to be a catalyst in their progress. We notably looked, defined and analyzed: what research is; what it requires; how it is engaged; how it is studied; and how it is measured.

## References

[1] P. D. Leedy, T. J. Newby and P. A. Ertmer, *Practical Research: Planning and Design,* 6[th] ed. Upper Saddle River, N.J.: Merrill, 1997, pp. 304.

[2] S. E. Matteson, "Common Threads in Research Across Disciplines: A Reflection," *First Annual Scholars Day Conference*, Denton, Texas, USA, Apr. 15, 2004.

[3] Y. R. Shayan and M. Abdulla, "Design Based Teaching for Science and Engineering Students," *5[th] Conference on Canadian Design Engineering Network (CDEN 2008)*, Halifax, Nova Scotia, Canada, Jul. 27-29, 2008.

[4] R. G. Gallager, "Low Density Parity Check Codes," *M.I.T. Press*, 1963.

[5] D. J.C. MacKay and R. M. Neal, "Near Shannon Limit Performance of Low Density Parity Check Codes," *Electronics Letters*, Jul. 1996.

[6] IEEE's Top 100 Documents. http://ieeexplore.ieee.org/Xplore/toparticles.jsp

[7] IEEE Communications Society (ComSoc). http://ww2.comsoc.org

[8] IEEE Xplore. http://ieeexplore.ieee.org

[9] ACM Portal. http://portal.acm.org

[10] SpringerLink. http://www.springerlink.com

[11] Elsevier ScienceDirect. http://www.sciencedirect.com

[12] SIAM Journals Online. http://epubs.siam.org

[13] ProQuest. http://www.proquest.com

[14] Concordia University's MetaFind Search Tool. http://library.concordia.ca/research/metafind

[15] COLOMBO's Inter-Library Loans Tool. http://library.concordia.ca/research/ill

[16] CREPUQ. http://www.crepuq.qc.ca

[17] R. Hamming, "You and Your Research," *Transcription of the Bell Communications Research Colloquium Seminar*, Morristown, New Jersey, USA, Mar. 7, 1986.

[18] E. Garfield, "The History and Meaning of the Journal Impact Factor," *Journal of the American Medical Association (JAMA)*, pp. 90-93, Jan. 2006.

[19] International Symposium on Peer Reviewing. http://www.ictconfer.org/ispr